\documentclass{aa}

\usepackage{amsmath, amssymb}
\usepackage[dvips]{graphicx}
\usepackage{epsfig}
\usepackage{color}
\usepackage{import}
\usepackage{diagbox}
\usepackage{natbib}

\usepackage{numprint}
\usepackage{siunitx}

\newcommand{\rot}{\mathbf{\nabla} \times}
\newcommand{\divg}{\mathbf{\nabla}\cdot}

\newcommand{\rlight}{r_{\rm L}}
\newcommand{\Rs}{R_{\rm s}}
\newcommand{\as}{a_{\rm s}}
\newcommand{\BQ}{B_{\rm Q}}

\newcommand{\ephi}{\mathbf{e}_\varphi}

\begin{document}

\title{Strongly magnetized rotating dipole in general relativity}


\author{J. P\'etri\thanks{E-mail: jerome.petri@astro.unistra.fr}}

\institute{Observatoire astronomique de Strasbourg, Universit\'e de Strasbourg, CNRS, UMR 7550, 11 rue de l'universit\'e, F-67000 Strasbourg, France.}

\date{Received / Accepted }

\abstract{Electromagnetic waves arise in many area of physics. Solutions are difficult to find in the general case.}{ In this paper, we numerically integrate Maxwell equations in a 3D spherical polar coordinate system.}{Straightforward finite difference methods would lead to a coordinate singularity along the polar axis. Spectral methods are better suited to deal with such artificial singularities related to the choice of a coordinate system. When the radiating object is rotating like for instance a star, special classes of solutions to Maxwell equations are worthwhile to study such as quasi-stationary regimes. Moreover, in high-energy astrophysics, strong gravitational and magnetic fields are present especially around rotating neutron stars.}{ In order to study such systems, we designed an algorithm to solve the time-dependent Maxwell equations in spherical polar coordinates including general relativity as well as quantum electrodynamical corrections to leading order. As a diagnostic, we compute the spindown luminosity expected from these stars and compare it to the classical i.e. non relativistic and non quantum mechanical results.}{It is shown that quantum electrodynamics leads to an irrelevant change in the spindown luminosity even for magnetic field around the critical value of $\numprint{4.4e9}$~\si{\tesla}. Therefore the braking index remains close to its value for a point dipole in vacuum namely $n=3$. The same conclusion holds for a general-relativistic quantum electrodynamically corrected force-free magnetosphere.
}

\keywords{gravitation - magnetic fields - plasmas - stars: neutron - methods: analytical - methods: numerical}

\maketitle

\section{Introduction}

Nature do not offer us much places in the Universe where to test our current theories of gravity and electromagnetism {\it simultaneously}. However and fortunately strong magnetic and gravitational fields exist inside and around neutron stars. They represent valuable laboratories to check our current theories in the strong field regime. Curvature of space-time is important because of the stellar compactness of about
\begin{equation}
\label{eq:compacite}
\Xi = \frac{\Rs}{R} \approx 0.345 \, \left( \frac{M}{1.4~M_\odot} \right) \, \left( \frac{R}{12 \textrm{ km}} \right)^{-1}
\end{equation}
where $R$ is the neutron star radius, $M$ its mass, $\Rs=2\,G\,M/c^2$ its Schwarzschild radius, $c$ the speed of light and $G$ the gravitational constant. Moreover, neutron stars are strongly magnetized objects, harbouring fields as high as the critical value of $\BQ\approx\numprint{4.4e9}$~\si{\tesla} or even higher. These regimes of strong gravity and magnetic fields are unreachable on Earth even separately. 

Since the exact analytical solution for a static dipole in general relativity (GR) found by \cite{Ginzburg1964} and those for multipolar terms in a spherically symmetric vacuum gravitational field by \cite{1972PhRvD...6.1476W}, several authors looked deeper into the effect of rotation with emphasizes to neutron stars. In vacuum, Maxwell equations remain linear even in a background gravitational field. This helped \cite{2001MNRAS.322..723R, 2002MNRAS.331..376Z, 2004MNRAS.352.1161R} to compute the electromagnetic field in the exterior of a slowly rotating neutron star. They gave approximate analytical expressions for the external electromagnetic field close to the neutron star which have later also been reported by \cite{2013MNRAS.433..986P}. \cite{2004MNRAS.348.1388K} extended the previous work by solving numerically the equations for the oblique rotator in vacuum in general relativity. They retrieve \cite{2001MNRAS.322..723R} results close to the surface and the Deutsch solution \citep{1955AnAp...18....1D} for distances larger than the light cylinder~$r\gg\rlight$ where $\rlight=c/\Omega$ and $\Omega$ is the rotation speed of the star.

Whereas quantum electrodynamics (QED) effects are known to be relevant for wave propagation in the birefringent vacuum as described by \cite{1971AnPhy..67..599A, 1979PhRvD..19.2868H, 1986ApJ...302..120A, 1988MNRAS.235...51B} or in the review by \cite{1992herm.book.....M, 2006RPPh...69.2631H, 2015SSRv..191...13L}, less attention has been focused so far to the whole picture of the magnetosphere. Let us mention the work of \cite{1997JPhA...30.6475H} who computed corrections to a dipole to first order for any strength of the magnetic field following \cite{1936ZPhy...98..714H} effective Lagrangian. This result has recently been generalized by \cite{2016MNRAS.456.4455P} taking into account the curvature of space-time following the 3+1~formalism developed by \cite{2015MNRAS.451.3581P}. We apply this formalism to a rotating monopole and dipole. Such corrections are relevant for magnetars, those neutron stars with the strongest magnetic fields known in the Universe \citep{2015RPPh...78k6901T}.

One of the mystery of the global electrodynamics of pulsar or neutron star magnetosphere is symbolized by the braking index~$n$ relating the braking torque to the rotation rate~$\Omega$ of the star by $\dot\Omega \propto - \Omega^n$ where a dot means derivative with respect to time. For a pure dipole rotating in vacuum, it should be very close to $n=3$, see for instance \cite{1973ApJ...181..161R} for the dipole and the general case of a multipole of order~$\ell$ being $n=2\,\ell+1$ as given in \cite{1991ApJ...373L..69K}, see also \cite{2015MNRAS.450..714P} for the exact expression taking into account the finite size of the star. In the dipole case \cite{2016MNRAS.455.3779P} showed that this result is not affected by the presence of a plasma in the magnetosphere in general relativity. \cite{2015PhRvD..91f3007H} summarize the state of the art in the measurements of pulsar braking indices. They are all less than~3, some of them much smaller, closer to 1 or 1.5 thus definitely ruling out a pure dipolar field in vacuum, force-free (FFE) or MHD regime. See however a very recent outsider reported by \cite{2016ApJ...819L..16A} to have $n=3.15$. Could QED effects account for this discrepancy? \cite{2012EL.....9849001D} claimed that QED can indeed strongly impact on the braking index. Starting from this assertion \cite{2016RAA....16a...9X} proposed to test the hypothesis of superstrong magnetic field in magnetars  by inspection of their energy loss that should be dominated by quantum vacuum friction. Even more recently \cite{2016ApJ...823...97C} build on this quantum vacuum friction effect and arrived at the same conclusion within a factor~2. Unfortunately as reported in this work, we do not retrieve their results. Their expression for QED corrections adds a spindown luminosity depending on $\Omega^2$ therefore resembles to radiation from a magnetic monopole very similar to the split monopole solution. Maxwell theory of electromagnetism does not allow radiation from a magnetic monopole in vacuum. Moreover, non linear electrodynamics of a rotating dipole would induce higher multipoles with mode numbers $\ell\geqslant1$ due to non-linearities but never a $\ell=0$ multipole. It is thus very difficult to understand the origin of the quantum vacuum luminosity given by these authors.

In this paper, we develop a pseudo-spectral discontinuous Galerkin method in space in the weak formulation to solve Maxwell equations in spherical coordinates using our formalism in general relativity with the effective Euler-Heisenberg QED Lagrangian. The set of equations and the solution techniques are reminded in Section~\ref{sec:GRElectrodynamic}. The algorithm is discussed in depth in Section~\ref{sec:Algorithm}. Results for the dipole in classical flat space-time and with strong field corrections from GR and QED are presented in Section~\ref{sec:ResultatsVide} for vacuum case and in Section~\ref{sec:ResultatsFFE} for FFE case. We conclude about possible extensions of this work in the concluding remarks of Section~\ref{sec:Conclusion}.

\section{Non linear electrodynamics in general relativity}
\label{sec:GRElectrodynamic}

In this section, we remind the equations satisfied by the electromagnetic field in general relativity, including quantum electrodynamical corrections following the 3+1 formalism detailed in \cite{2015MNRAS.451.3581P}. These equations are then written in component form introducing contravariant and covariant components for the electromagnetic tensor and related fields. Eventually, we explain how to solve this system numerically.

\subsection{The field equations}

In a 3+1 foliation of spacetime, the equations for the electromagnetic field are very similar to their flat spacetime counterpart. Indeed, Maxwell equations taking into account general-relativistic as well as quantum electrodynamical corrections are given by
\begin{subequations}
\label{eq:GRQEDMaxwell}
\begin{align}
  \label{eq:Maxwell_Faraday}
  \rot \mathbf{E} & = - \frac{1}{\sqrt{\gamma}} \, \partial_t (\sqrt{\gamma} \, \mathbf{B} ) \\
  \label{eq:Maxwell_Ampere}
 \rot \mathbf{H} & = \mathbf{J} + \frac{1}{\sqrt{\gamma}} \, \partial_t (\sqrt{\gamma} \, \mathbf{D})
 \end{align}
supplemented with the initial condition on the divergence
\begin{align}
\label{eq:Maxwell_Div_B}
  \divg \mathbf{B} & = 0 \\
\label{eq:Maxwell_Div_D}
 \divg \mathbf{D} & = \rho \;.
\end{align}
\end{subequations}
$\gamma$ represents the determinant of the spatial metric, $\mathbf{J}$ the current density and $(\mathbf{E}, \mathbf{B}, \mathbf{D}, \mathbf{H})$ the various representative electromagnetic fields. In order to include QED effects, we furthermore introduce two auxiliary vector fields denoted by $\mathbf{F}$ and $\mathbf{G}$. One set of constitutive relations is derived from the 3+1 decomposition of space-time and reads
\begin{subequations}
\label{eq:Constitutive}
\begin{align}
\label{eq:ConstitutiveE}
  \varepsilon_0 \, \mathbf E & = \alpha \, \mathbf F + \varepsilon_0\,c\,\pmb{\beta} \times \mathbf B \\
\label{eq:ConstitutiveG}
  \mu_0 \, \mathbf G & = \alpha \, \mathbf B - \frac{\mathbf\beta \times \mathbf 
F}{\varepsilon_0\,c} \;.
\end{align}
\end{subequations}
The space-time geometry is described by the lapse function~$\alpha$, the shift vector~$\pmb \beta$ and the spatial metric~$\gamma_{ab}$ whose determinant is $\gamma$. The other set is derived from the Euler-Heisenberg Lagrangian and given by introducing two parameters $(\xi_1, \xi_2)$ such that
\begin{subequations}
\label{eq:FGDBEH}
 \begin{align}
\label{eq:DFB}
 \mathbf{D} & = \xi_1 \, \mathbf{F} + \frac{\xi_2}{c} \, \mathbf{B} \\
\label{eq:HEG}
 \mathbf{H} & = \xi_1 \, \mathbf{G} - \frac{\xi_2}{c} \, \mathbf{E} \;.
 \end{align}
\end{subequations}
From the first order perturbation of the Lagrangian of the electromagnetic field, we found that these two parameters are given by
\begin{subequations}
 \begin{align}
 \xi_1 & = 1 - 16\,\mu_0\,\eta_1 \, \left( B^2 - \frac{\mu_0}{\varepsilon_0} \, F^2 \right) \\
 \xi_2 & = 32 \, \eta_2 \, \frac{\mathbf{F} \cdot \mathbf{B}}{\varepsilon_0\,c} 
\;.
 \end{align}
\end{subequations}
In the Euler-Heisenberg prescription we have immediately that
\begin{subequations}
\begin{align}
 \eta_1 & = \frac{\alpha_{\rm sf}}{180\,\pi} \, \frac{1}{2\,\mu_0\,\BQ^2} \\
 \eta_2 & = \frac{7}{4} \, \eta_1 
\end{align}
\end{subequations}
with $\alpha_{\rm sf}$ the fine structure constant and $\BQ\approx\numprint{4.4e9}$~\si{\tesla} the critical magnetic field strength. Note that the perturbations to Maxwell equations are treated as done in gravitational theory by using post-Newtonian expansion requiring several parameters. In any case the corrections remain small that is $(\eta_1,\eta_2)\ll1$. It would therefore in principle be possible to use the Born-Infeld Lagrangian in the weak field limit if we set 
\begin{subequations}
\begin{align}
 \eta_1 & = \frac{1}{32\,\mu_0\,b^2} \\
 \eta_2 & = \eta_1
\end{align}
\end{subequations}
with $b=\numprint{9.18e11}$~\si{\tesla} the empirical maximal absolute field strength in Born-Infeld theory. To summarize we have six vector fields $(\mathbf{F}, \mathbf{B}, \mathbf{E}, \mathbf{H}, \mathbf{D}, \mathbf{G})$ satisfying two evolution equations (\ref{eq:Maxwell_Faraday}),(\ref{eq:Maxwell_Ampere}) two constraints (\ref{eq:Maxwell_Div_B}),(\ref{eq:Maxwell_Div_D}) and four constitutive relations eq.(\ref{eq:ConstitutiveE}), (\ref{eq:ConstitutiveG}), (\ref{eq:DFB}) and (\ref{eq:HEG}) .



\subsection{Field equations in component form}

In order to deal with any kind of curvilinear coordinate system, we write the field equations in component form adapted to an absolute space~$x^a$ and a time coordinate~$t$ as described by an observer with four velocity~$n^i$, indices $a$ to $h$ span the spatial part whereas indices starting from $i$ span the four-dimensional space-time. The time evolution of the electric and magnetic fields $\mathbf{D}$ and $\mathbf{B}$ is therefore given by
\begin{subequations}
 \begin{align}
  \partial_t ( \sqrt{\gamma} \, D^a ) & =   \varepsilon^{abc} \, \partial_b H_c \\
  \partial_t ( \sqrt{\gamma} \, B^a ) & = - \varepsilon^{abc} \, \partial_b E_c
 \end{align}
\end{subequations}
with the constitutive relations expressed to first order in the QED parameters
\begin{subequations}
 \begin{align}
  \varepsilon_0 \, E_a & = \frac{\alpha}{\xi_1} \, D_a + \varepsilon_0 \, c \, \sqrt{\gamma} \, \varepsilon_{abc} \, \beta^b \, B^c - \frac{\alpha\,\xi_2}{c} \, B_a \\
  \mu_0 \, H_a & = \alpha \, \xi_1 \, B_a - \sqrt{\gamma} \, \varepsilon_{abc} \, \frac{\beta^b \, D^c}{\varepsilon_0 \, c} - \frac{\alpha\,\xi_2\,\mu_0}{\varepsilon_0\,c} \, D_a
 \end{align}
\end{subequations}
and the constraint equations
\begin{subequations}
 \begin{align}
  \frac{1}{\sqrt{\gamma}} \, \partial_a ( \sqrt{\gamma} \, D^a ) & = 0 \\
  \frac{1}{\sqrt{\gamma}} \, \partial_a ( \sqrt{\gamma} \, B^a ) & = 0 \; .
 \end{align}
\end{subequations}
In the slow rotation approximation frequently used for neutron stars, the metric is essentially described by two parameters: the Schwarzschild radius defined by
\begin{equation}
 \Rs = \frac{2\,G\,M}{c^2}
\end{equation}
and the spin parameter~$\as$ which is left as a free quantity. A reasonable choice for spherically symmetric neutron stars would be
\begin{equation}
\label{eq:SpinParameter}
 \frac{\as}{\Rs} = \frac{2}{5} \, \frac{R}{\Rs} \, \frac{R}{\rlight} \; .
\end{equation}
The spatial metric is given in spherical Boyer-Lindquist coordinates by
\begin{equation}
  \label{eq:Metric3D}
  \gamma_{ab} =
  \begin{pmatrix}
    \alpha^{-2} & 0 & 0 \\
    0 & r^2 & 0 \\
    0 & 0 & r^2 \sin^2\vartheta
  \end{pmatrix}
\end{equation}
where the lapse function is
\begin{equation}
  \label{eq:Lapse}
  \alpha = \sqrt{ 1 - \frac{\Rs}{r} }
\end{equation}
and the shift vector
\begin{subequations}
 \begin{align}
  \label{eq:Shift}
  c \, \beta = & - \omega \, r \, \sin\vartheta \, \ephi \\
  \omega = & \frac{\as\,\Rs\,c}{r^3}
 \end{align}
\end{subequations}
and in contravariant components the only non vanishing term is simply $\beta^\varphi = - \omega/c$.
See \cite{2013MNRAS.433..986P, 2014MNRAS.439.1071P,2015MNRAS.447.3170P} for more details about the 3+1~foliation.

\subsection{Vacuum polarization}

Electrodynamics in the presence of strong electromagnetic fields can be described by the above non linear Maxwell equations derived from an effective Lagrangian computed in the limit $B\ll \BQ$ by Euler and Heisenberg. Quantum electrodynamics describes vacuum as a polarized and magnetized media without external current density~$\mathbf{J}=\mathbf{0}$ or charge density~$\rho=0$. 

The usual convention in special relativity introduces the vector fields $(\mathbf{D}, \mathbf{H})$ according to the first order expansion in the fine structure constant, as given for example by Euler and Heisenberg Lagrangian by
\begin{subequations}
 \begin{align}
  \mathbf D & = \varepsilon_0 \, \mathbf E + \kappa \, ( 2 \, ( E^2 - c^2\,B^2 ) \, \mathbf E + 7 \, c^2 \, (\mathbf{E} \cdot \mathbf{B}) \, \mathbf B ) \\
  \mathbf H & = \frac{\mathbf B}{\mu_0} + \kappa \,  ( 2 \, c^2\,( E^2 - c^2\,B^2 ) \, \mathbf B - 7 \, c^2\,(\mathbf{E} \cdot \mathbf{B}) \, \mathbf E )
 \end{align}
\end{subequations}
with
\begin{equation}
 \kappa = \frac{\alpha_{\rm sf}}{45\,\pi\,\mu_0\,c^4\,B_q^2} \ .
\end{equation}
If we include the effect of a gravitational field, in our new convention, the vector fields $(\mathbf{F}, \mathbf{G})$ must be understood as being the vector fields $(\mathbf{D}, \mathbf{H})$ as seen in eq.~(\ref{eq:FGDBEH}) whereas $(\mathbf{F}, \mathbf{B})$ are the fields measured by a local observer.

The field eq.~(\ref{eq:GRQEDMaxwell}) evolve the primary vectors $\mathbf{B}$ and $\mathbf{D}$. The other auxiliary fields are deduced from the four constitutive relations. The implementation of the numerical algorithm is as follows. To advance all the quantities one time step into the future, let us assume that the fields $\mathbf{B}$ and  $\mathbf{D}$ are known at the initial stage. Then $\mathbf{F}$ can be retrieved from eq.~(\ref{eq:DFB}). Next from the knowledge of $\mathbf{F}$ and $\mathbf{B}$ the fields $\mathbf{E}$ and $\mathbf{G}$ are retrieved through eq.~(\ref{eq:Constitutive}). Finally $\mathbf{H}$ is obtained from eq.~(\ref{eq:HEG}) knowing $\mathbf{E}$ and $\mathbf{G}$ from the previous calculation. This completes one full time step to advance the primary fields $\mathbf{B}$ and $\mathbf{D}$. The constitutive relations from general relativity are linear such that it is straightforward to compute the two unknown fields from the two known fields. The complication arises from the vacuum polarization relations because they are non linear. Getting $\mathbf{F}$ from $\mathbf{B}$ and $\mathbf{D}$ would be difficult because eq.~(\ref{eq:DFB}) is non linear in the unknown $\mathbf{F}$ because of the parameter~$\xi_1$. Nevertheless, as our equations are valid only up to first order in the fine structure constant, it is sufficient to invert this relation to the same order of accuracy. Therefore, we only need to plug $\mathbf{D}$ into the parameters~$\xi_1$ and $\xi_2$ instead of $\mathbf{F}$. This trick enables us to compute straightforwardly the auxiliary fields without resorting to an inversion of a non linear system.

For numerical purposes, for the remainder of this paper, we normalize electromagnetic quantities with respect to the critical field $\BQ$ and its derivatives like $c\,\BQ$, $\varepsilon_0\,c\,\BQ$, $\BQ/\mu_0$ for the other fields and use units with $c = \varepsilon_0 = \mu_0 = 1$. Denoting these fields with lower cases, the normalized system to be solved reads
\begin{subequations}
\label{eq:Maxwell_Normees}
 \begin{align}
 \frac{1}{\sqrt{\gamma}} \, \partial_t (\sqrt{\gamma} \, \mathbf{b} ) & = - \rot \mathbf{e} \\
 \frac{1}{\sqrt{\gamma}} \, \partial_t (\sqrt{\gamma} \, \mathbf{d}) & = \rot \mathbf{h} \\
 \xi_1 \, \mathbf{f} & = \mathbf{d} - \xi_2 \, \mathbf{b} \\
 \mathbf e & = \alpha \, \mathbf f + \mathbf \beta \times \mathbf b \\
 \mathbf g & = \alpha \, \mathbf b - \mathbf \beta \times \mathbf f \\
 \mathbf{h} & = \xi_1 \, \mathbf{g} - \xi_2 \, \mathbf{e}
 \end{align}
with the normalized parameters in Euler-Heisenberg QED to first order
\begin{align}
 \xi_1 & = 1 + \frac{2\,\alpha_{\rm sf}}{45\,\pi} \, \left( d^2 - b^2 \right) \\
 \xi_2 & = \frac{7\,\alpha_{\rm sf}}{45\,\pi} \, \mathbf{d} \cdot \mathbf{b} \;.
\end{align}
\end{subequations}
Note that we replaced $\mathbf{f}$ by $\mathbf{d}$ as previously discussed.

\section{Algorithm}
\label{sec:Algorithm}

\subsection{Boundary conditions}

As in \cite{2014MNRAS.439.1071P} and in \cite{2016MNRAS.455.3779P} we put boundary conditions on the neutron star surface according to the magnetic frozen in assumption. In the most general regime, including gravitation and vacuum polarization, the jump conditions at the stellar surface still enforce continuity of the normal component of the magnetic field~$B^{r}$ and continuity of the tangential component of the electric field~$\{E^{\vartheta}, E^{\varphi}\}$. More explicitly, they are such that
\begin{subequations}
  \label{eq:CLimites}
\begin{align}
  B^{r}(t,R,\vartheta,\varphi) & = B^{r}_0(t,\vartheta,\varphi) \\
  D^{\vartheta}(t,R,\vartheta,\varphi) & = \frac{\xi_2}{c} \, B^\vartheta - \varepsilon_0 \, \xi_1 \, \frac{\Omega-\omega}{\alpha^2} \, \sin\vartheta \, B^{r}_0(t,\vartheta,\varphi) \\
  D^{\varphi}(t,R,\vartheta,\varphi) & = \frac{\xi_2}{c} \, B^\varphi \;.
\end{align}
\end{subequations}
However, the stationary magnetic field $B^{r}_0$ contains corrections due to QED. We treat the problem to first order in the perturbation of the Lagrangian but to any order in the compactness. Approximate analytical solutions for a strongly magnetized oblique dipole in general relativity are given by \cite{2016MNRAS.456.4455P}. We use these expressions for the stellar interior.

The continuity of $B^{r}$ automatically implies the correct boundary treatment of the electric field. $B^{r}_0(t,\vartheta,\varphi)$ represents the, possibly time-dependent, radial magnetic field imposed by the star, let it be monopole, split monopole, oblique dipole or multipole.

Maxwell equations in QED vacuum are
\begin{subequations}
 \begin{align}
  \partial_t ( \sqrt{\gamma} \, D^r ) & = \partial_\vartheta H_\varphi - \partial_\varphi H_\vartheta \\
  \partial_t ( \sqrt{\gamma} \, D^\vartheta ) & = \partial_\varphi H_r - \partial_r H_\varphi \\
  \partial_t ( \sqrt{\gamma} \, D^\varphi ) & = \partial_r H_\vartheta - \partial_\vartheta H_r \\
  \partial_t ( \sqrt{\gamma} \, B^r ) & = \partial_\varphi E_\vartheta - \partial_\vartheta E_\varphi \\
  \partial_t ( \sqrt{\gamma} \, B^\vartheta ) & = \partial_r E_\varphi - \partial_\varphi E_r \\
  \partial_t ( \sqrt{\gamma} \, B^\varphi ) & = \partial_\vartheta E_r - \partial_r E_\vartheta \;.
 \end{align}
\end{subequations}
We look for the characteristics propagating along the radial direction. To this end, we isolate expressions containing the radial propagation that is $\partial_r$ and $\partial_t$. Eliminating all useless terms for this radial propagation, the system reduces to
\begin{subequations}
 \begin{align}
  \partial_t ( \sqrt{\gamma} \, D^\vartheta ) + \partial_r H_\varphi & = 0 \\
  \partial_t ( \sqrt{\gamma} \, D^\varphi ) - \partial_r H_\vartheta & = 0 \\
  \partial_t ( \sqrt{\gamma} \, B^\vartheta ) - \partial_r E_\varphi & = 0 \\
  \partial_t ( \sqrt{\gamma} \, B^\varphi ) + \partial_r E_\vartheta & = 0 \;.
 \end{align}
\end{subequations}
The covariant components of the spatial vectors $\mathbf{D}$ and $\mathbf{B}$ are giving by lowering the indexes such that for a diagonal spatial metric given by eq.~(\ref{eq:Metric3D}) we have
\begin{subequations}
 \begin{align}
 D_\vartheta & = \gamma_{\vartheta \vartheta} \, D^\vartheta \\
 D_\varphi & = \gamma_{\varphi \varphi} \, D^\varphi \\
 B_\vartheta & = \gamma_{\vartheta \vartheta} \, B^\vartheta \\
 B_\varphi & = \gamma_{\varphi \varphi} \, B^\varphi .
 \end{align}
\end{subequations}
Injecting the constitutive relations into the evolution equations
and defining the unknown vector
\begin{equation}
U =
 \begin{pmatrix}
  \sqrt{\gamma} \, \mu_0 \, D^\vartheta \\
  \sqrt{\gamma} \, \mu_0 \, D^\varphi \\
  \sqrt{\gamma} \, \varepsilon_0 \, B^\vartheta \\
  \sqrt{\gamma} \, \varepsilon_0 \, B^\varphi
 \end{pmatrix}
\end{equation}
the system can be cast into the conservative form $\partial_t U + \partial_r (A\,U) = 0$ with
\begin{equation}
A =
 \begin{pmatrix}
  -c\,\beta^r & -\frac{\alpha^2\,\xi_2\sin\vartheta}{\varepsilon_0\,c} & 0 & \frac{\alpha^2\,\xi_1\sin\vartheta}{c} \\
  \frac{\alpha^2\,\xi_2\sin\vartheta}{\varepsilon_0\,c} & -c\,\beta^r & -\frac{\alpha^2\,\xi_1\sin\vartheta}{c} &  0 \\
  0 & -\frac{\alpha^2\,\sin\vartheta}{\mu_0\,\xi_1}  & -c\,\beta^r & \frac{\alpha^2\,\xi_2\,\sin\vartheta}{\varepsilon_0\,c} \\
  \frac{\alpha^2}{\mu_0\,\xi_1\,\sin\vartheta} & 0 & -\frac{\alpha^2\,\xi_2}{\varepsilon_0\,c\,\sin\vartheta} & -c\,\beta^r
 \end{pmatrix}.
\end{equation}
For the slowly rotating metric in spherical Boyer-Lindquist coordinates these expressions simplify. The eigenvalues for the electromagnetic waves propagating in the QED vacuum in general relativity are given to first order in the parameters $(\xi_1,\xi_2)$ by
\begin{equation}
 ( - \beta^r \pm \alpha^2 ) \, c
\end{equation}
and the eigenvectors by
\begin{subequations}
 \begin{align}
  ( \pm \frac{\xi_1\,\sin\vartheta}{\varepsilon_0\,c}, \frac{\xi_1\,\xi_2}{\varepsilon_0^2\,c^3}, 0, 1) & \ \ \ ( \frac{\xi_1\,\xi_2}{\varepsilon_0^2\,c^3}, \pm \frac{\xi_1}{\varepsilon_0\,c\,\sin\vartheta}, 1, 0) .
 \end{align}
\end{subequations}
The characteristics that propagate are
\begin{subequations}
 \begin{align}
  \varepsilon_0\,c\,\xi_1\,\sin\vartheta\, B^\varphi & \pm ( D^\vartheta - \frac{\xi_1\,\xi_2}{c} \, B^\vartheta ) \\
  \varepsilon_0\,c\,\xi_1\,B^\vartheta & \pm \sin\vartheta \, ( D^\varphi - \frac{\xi_1\,\xi_2}{c} \, B^\varphi ) \;.
 \end{align}
\end{subequations}
The outer boundary condition cannot be handled exactly. We need to make some approximate assumptions about the outgoing waves we want to enforce in order to prevent reflections from this artificial outer boundary. Using the Characteristic Compatibility Method (CCM) described in \cite{Canuto2007} and neglecting frame-dragging and strong field effects far from the neutron star, the radially propagating characteristics are given to good accuracy by their flat space-time counterpart as
\begin{eqnarray}
  \label{eq:CCM1}
  D^{\vartheta} \pm \varepsilon_0 \, c\, \sin\vartheta \, B^{\varphi} & ; & \sin\vartheta \, D^{\varphi} \pm \varepsilon_0 \, c\, B^{\vartheta}.
\end{eqnarray}
In order to forbid ingoing wave we ensure that
\begin{subequations}
\begin{align}
  \label{eq:CCM2}
  D^{\vartheta} - \varepsilon_0 \, c\, \sin\vartheta \, B^{\varphi} & = 0 \\
  \label{eq:CCM3}
  \sin\vartheta \, D^{\varphi} + \varepsilon_0 \, c\, B^{\vartheta} & = 0
\end{align}
whereas the other two characteristics are found by
\begin{align}
  \label{eq:CCM4}
  D^{\vartheta} + \varepsilon_0 \, c\, \sin\vartheta \, B^{\varphi} & = D^{\vartheta}_{\rm PDE} + \varepsilon_0 \, c\, \sin\vartheta \, B^{\varphi}_{\rm PDE} \\
  \label{eq:CCM5}
  \sin\vartheta \, D^{\varphi} - \varepsilon_0 \, c\, B^{\vartheta} & = \sin\vartheta \, D^{\varphi}_{\rm PDE} - \varepsilon_0 \, c\, B^{\vartheta}_{\rm PDE}
\end{align}
\end{subequations}
the subscript $_{\rm PDE}$ denoting the values of the electromagnetic field obtained by straightforward time advancing without care of any boundary condition. The new corrected values are deduced from the solution of the linear system made of equations~(\ref{eq:CCM2})-(\ref{eq:CCM5}).

%
%
%
%

%
%

\section{Vacuum results}
\label{sec:ResultatsVide}

We apply our new code to some typical magnetic field topologies such as a pure monopole and a pure dipole magnetic field taking into account GR and QED. Four different regimes are investigated corresponding to Newtonian or general-relativistic and quantum or classical (in the quantum sense) approximation. As a representative sample of these four approaches, we use typical parameters summarized in table~\ref{tab:Parametres}.
\begin{table}
\centering
\begin{tabular}{c|cc}
 & classical & quantum \\
\cline{2-3}
Newtonian (N) & $(0.0,-3)$ & $(0.0,0)$ \\
General-relativistic (GR) & $(0.5,-3)$ & $(0.5,0)$ \\
\cline{2-3}
\end{tabular}
\caption{The two parameters $(\Rs/R,\log(B/\BQ))$ describing the actual regime investigated.}
\label{tab:Parametres}
\end{table}
We start with axisymmetric fields and then discuss about the orthogonal and oblique dipole rotator. If not specified otherwise, we use a five points Legendre interpolation scheme for vacuum fields and a three points scheme (quadratic polynomials) for FFE fields in each cell in the radial direction. The numerical resolution for vacuum fields is $N_{\rm r} \times N_\vartheta \times N_\varphi = 128 \times 8 \times 16$ and for FFE fields we use $N_{\rm r} \times N_\vartheta \times N_\varphi = 64 \times 32 \times 64$.  We also introduce the normalized rotation rate as $a=R/\rlight$.

\subsection{Vacuum monopole}

\begin{figure}
\centering
\input{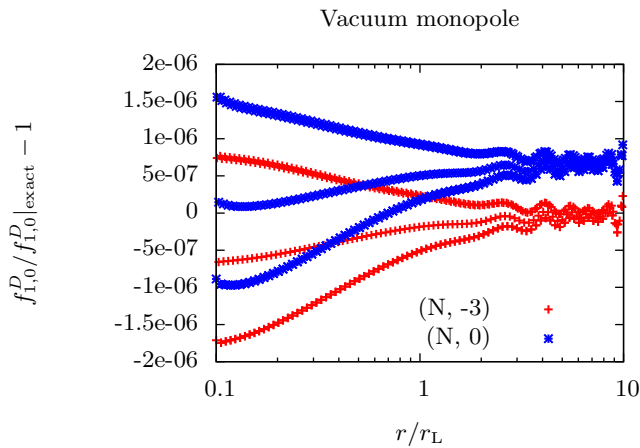}
\caption{Electric field potential $f^D_{1,0}$ of the vacuum monopole field for $a=0.1$, $b=10^{-3}$ in red and $b=1$ in blue, for Newtonian gravity.}
\label{fig:Monopole_f_D_lm_newt}
\end{figure}

For the monopole magnetic field, we compare the electric field components found from the simulations to those obtained for the classical Newtonian (in the sense of non relativistic and non quantum mechanical) rotating monopole. The non vanishing covariant components of the electric field are represented by the potential
\begin{equation}
 f^D_{1,0}(r) = 2 \, \sqrt{\frac{2\,\pi}{3}} \, \frac{\Omega\,B\,R^4}{r^2} \ .
\end{equation}
This function is compared to the simulations in fig.~\ref{fig:Monopole_f_D_lm_newt}.
The analytical results is retrieved to very good accuracy, more than 5~digits in classical but also in the quantum case. Actually both results are undistinguishable, they overlap perfectly. We conclude that QED has little impact on the monopolar electromagnetic field structure and can be neglected. This can directly be derived from the weakness of the correcting factors $(\xi_1,\xi_2)$ which bring perturbations to Maxwell equations of the order of several times
\begin{equation}
 \frac{\alpha_{\rm sf}}{45\,\pi} \, b^2 \approx 5 \times 10^{-5} \, b^2 \ll b^2 \ .
\end{equation}
QED corrections do not have a significant impact on the global topology of the magnetosphere.

\begin{figure}
\centering
\input{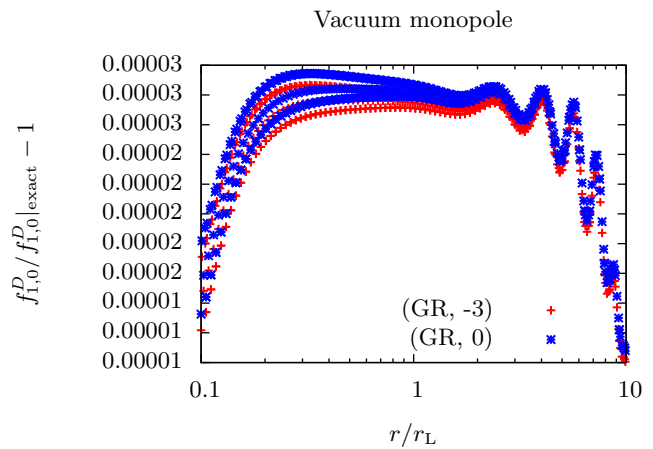}
\caption{Electric field potential $f^D_{1,0}$ of the vacuum monopole field for $a=0.1$, $b=10^{-3}$ in red and $b=1$ in blue, for general relativity.}
\label{fig:Monopole_f_D_lm_GR}
\end{figure}

The same observations apply to the general-relativistic fields. Indeed, some approximate solutions are known to first order in $\Rs$ and given by eq.~(27) in \cite{2015MNRAS.447.3170P}. We compare this analytical solution for $f^D_{1,0}$ to the output of our simulations in fig.~\ref{fig:Monopole_f_D_lm_GR}. Both simulation results overlap and agree with the approximate analytical solution. GR leads to much stronger perturbations of the electromagnetic field compared to QED. Here also we can neglect its influence.

\subsection{Vacuum aligned dipole}

\begin{figure}
\centering
\input{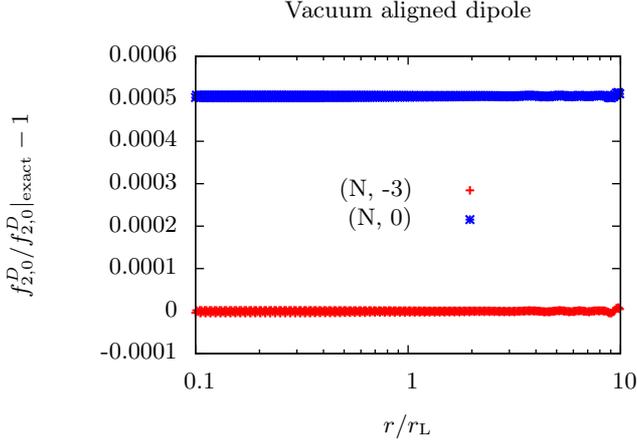}
\caption{Electric field potential $f^D_{2,0}$ of the vacuum dipole field for $a=0.1$, $b=10^{-3}$ in red and $b=1$ in blue, in Newtonian gravity.}
\label{fig:Dipole_f_D_lm_newt}
\end{figure}

The same study is performed for the rotating dipole. We compare again the electric field components from the simulation to those obtained for the classical and general-relativistic dipole. The non vanishing components in Newtonian gravity are derived from the potential
\begin{equation}
 f^E_{2,0}(r) = \sqrt{\frac{8\,\pi}{15}} \, \frac{\Omega\,B\,R^5}{r^3} \ .
\end{equation}
This function is compared to the simulations in fig.~\ref{fig:Dipole_f_D_lm_newt}.
The analytical results is retrieved to very good accuracy, more than 3 digits, and also in the quantum case. Actually both results are undistinguishable, they overlap perfectly. We conclude that QED has little impact on the dipolar electromagnetic field structure and can be neglected.

\begin{figure}
\centering
\input{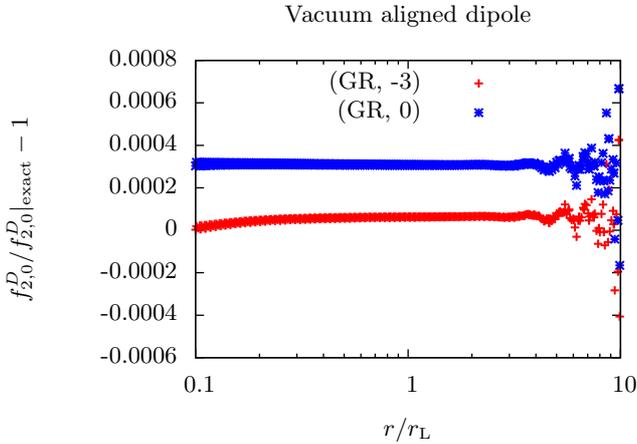}
\caption{Electric field potential $f^D_{2,0}$ of the vacuum dipole field for $a=0.1$, $b=10^{-3}$ in red and $b=1$ in blue, in general relativity.}
\label{fig:Dipole_f_D_lm_GR}
\end{figure}

The same observations apply to the general-relativistic fields where comparisons are made possible thanks to some approximate solutions given to first order in $\Rs$ by eq.~(55) in \cite{2013MNRAS.433..986P}. We compare this analytical solution for $f^D_{2,0}$ to the output of our simulations in fig.~\ref{fig:Dipole_f_D_lm_GR}. Both simulation results overlap and agree with the approximate analytical solution. GR leads to much stronger perturbations of the electromagnetic field compared to QED as in the monopole field. Here again QED can be ignored.

\subsection{Vacuum orthogonal dipole}

\begin{figure}
\centering
\input{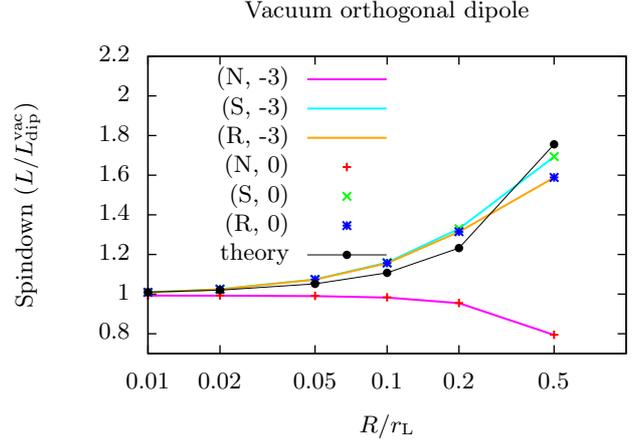}
\caption{Spindown luminosity for the vacuum orthogonal rotator for different rotation rates~$R/\rlight=\{0.01,0.02,0.05,0.1,0.2,0.5\}$, magnetic field strengths given by $\log(b)$ and gravitational field (Newtonian or GR) as indicated in the legend.}
\label{fig:Spindown_Dipole}
\end{figure}

Let us continue with the most interesting case, the orthogonal rotator which emits strong amplitude low frequency electromagnetic waves in vacuum. As a diagnostic, we compute the Poynting flux, i.e. its spindown luminosity depending on the two parameters presented in table~\ref{tab:Parametres}. The typical spin-down luminosity, used for normalization, is given by the classical flat space-time orthogonal rotator
\begin{equation}
\label{eq:SpinDownDipole}
L_{\rm dip}^{\rm vac} = \frac{8\,\pi}{3} \, \frac{\Omega^4\,B^2\,R^6}{\mu_0\,c^3} \ .
\end{equation}
We also investigated the influence of the neutron spin depicted by the ratio $R/\rlight$. Results are shown in fig.~\ref{fig:Spindown_Dipole}. We distinguish three different gravity regimes. The first one is Newtonian gravity thus flat space-time shown as N (for Newtonian) in the legend, the second is a Schwarzschild metric not including frame-dragging effects depicted by S (for Schwarzschild) in the legend and a third full GR regime with lapse function different from unity and non-vanishing shift vector denoted by R (for rotating). There is no distinction between classical and QED spindown luminosity. Both cases overlap again to high accuracy. However, to better assess the discrepancy between both regime, the explicit value of the Poynting fluxes are reported in table~\ref{tab:Spindown}. Values agree within 3 to 4 digits.
\begin{table}
\centering
\begin{center}
\begin{tabular}{ccccc}
\hline 
 & \multicolumn{2}{c}{Newtonian} & \multicolumn{2}{c}{General-relativistic} \\
\hline 
\diagbox{$a$}{$\log b$} & $-3$ & $0$ & $-3$ & $0$ \\
\hline 
0.01 & 0.9924 & 0.9933 & 1.0083 & 1.0089 \\
0.02 & 0.9921 & 0.9930 & 1.0244 & 1.0249 \\
0.05 & 0.9900 & 0.9909 & 1.0733 & 1.0739 \\
0.1  & 0.9826 & 0.9835 & 1.1556 & 1.1562 \\
0.2  & 0.9542 & 0.9551 & 1.3139 & 1.3146 \\
0.5  & 0.7941 & 0.7948 & 1.5875 & 1.5884 \\
\hline
\end{tabular}
\end{center}
\caption{Spindown luminosity for the vacuum orthogonal rotator in several approximations.}
\label{tab:Spindown}
\end{table}
In all cases, the spindown shows a small dependence on the spin rate through the ratio $R/\rlight$. In the often quoted point dipole limit, $R=0$ and such dependence would disappear. However, when the finite size of the star is taken into account, electric charges and currents built on the stellar surface and exert an additional torque on the star. Moreover this charge distribution induces an electric quadrupolar field that contributes to the overall electromagnetic radiation and spindown losses. These electric corrections add terms proportional to powers of $a$ thus explaining the variation of $L_{\rm dip}^{\rm vac}$ with spin frequency. In the plot, we also recognize an opposite slope in the dependence on the spin, negative for Newtonian gravity and positive for GR. The negative slope of Newtonian gravity is reminiscent of the lowest order corrections given by $L^{\rm vac} \approx (1-a^2) \, L_{\rm dip}^{\rm vac}$. The increase is spindown luminosity in the general-relativistic case can partially be attributed to the increase in the strength of the transverse radiating magnetic field~$B_{\rm T}$ at the light cylinder. The ratio~$(B_{\rm T}^{\rm GR}/B_{\rm T}^{\rm N})^2$ is shown in solid black line with dots (see \cite{2004MNRAS.352.1161R} for another estimate) and denoted theory in the legend. The upward trend is clearly apparent although it does not account for the full increase. The case $a=0.5$ is pathological because stellar boundary conditions strongly perturb these simple estimates. If frame-dragging is included, the actual rotation rate of the neutron star as measured by a local observer is reduced to a rate of $\Omega-\omega$ thus decreasing the electric field at the surface. Consequently the spindown luminosity, proportional to a power of $\Omega$ in flat spacetime, is also slightly decreased as seen in fig.~\ref{fig:Spindown_Dipole}, compare the ``R'' cases to the ``S'' cases.

\subsection{Vacuum oblique dipole}

\begin{figure}
\centering
\input{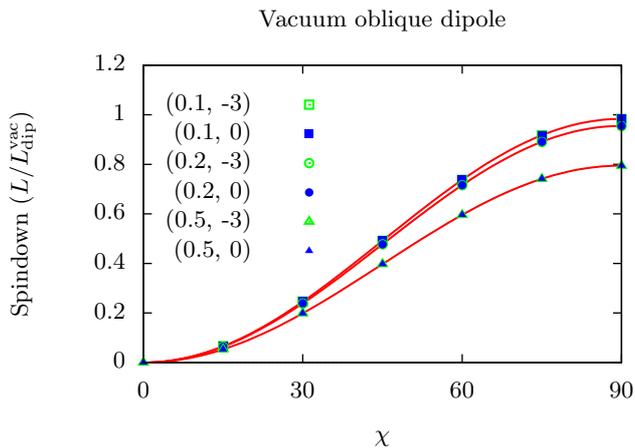}
\caption{Spindown luminosity for oblique rotators for different rotation rates and magnetic field strengths given by the couple $(a,\log(b))$ as indicated in the legend, for Newtonian gravity. Red solid lines are best fits.}
\label{fig:Spindown_Dipole_Oblique_N}
\end{figure}

\begin{figure}
\centering
\input{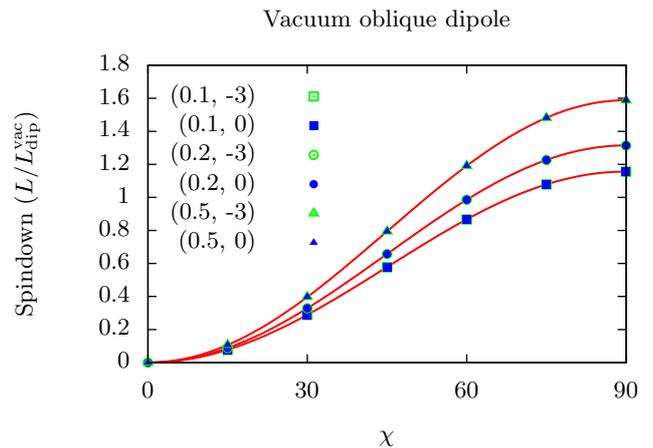}
\caption{Spindown luminosity for oblique rotators for different rotation rates and magnetic field strengths given by the couple $(a,\log(b))$ as indicated in the legend, in general relativity. Red solid lines are best fits.}
\label{fig:Spindown_Dipole_Oblique_GR}
\end{figure}

We finish with the oblique rotator by estimating the dependence of the spindown luminosity on the inclination angle~$\chi$ of the dipole. The Poynting flux with respect to $a$, $\log(b)$ and $\chi$ is shown in fig.~\ref{fig:Spindown_Dipole_Oblique_N} for Newtonian gravity and in fig.~\ref{fig:Spindown_Dipole_Oblique_GR} for general relativity. Here again we do not notice any significant deviation from the classical approximation neither in the Newtonian regime nor in general relativity.
The points are taken from the simulations whereas the solid curves are best fits obtained by adjusting to a $\sin^2\chi$ dependence such that
\begin{equation}
 \frac{L}{L_{\rm dip}^{\rm vac}} = \mathcal{L}^{\rm vac}_\perp \, \sin^2\chi .
\end{equation} 
The precise values are reported in Table~\ref{tab:FitFluxVacuum} comparing Newtonian and general-relativistic gravity. As the ratio~$R/\rlight$ decreases, both kind of curves, Newtonian and general-relativistic, tend to the function~$\sin^2\chi$, the former from below and the latter from above which means $\lim_{a \rightarrow 0} \mathcal{L}^{\rm vac}_\perp = 1$.
\begin{table}
\centering
\begin{center}
\begin{tabular}{ccccc}
\hline
 & \multicolumn{2}{c}{Newtonian} & \multicolumn{2}{c}{GR} \\
 \hline
\diagbox{$a$}{$\log b$} & $-3$ & $0$ & $-3$ & $0$ \\
\hline
\hline
0.1 & 0.9826 & 0.9835 & 1.1556 & 1.1562 \\
0.2 & 0.9542 & 0.9551 & 1.3139 & 1.3146 \\
0.5 & 0.7941 & 0.7948 & 1.5875 & 1.5884 \\
\hline
\end{tabular}
\end{center}
\caption{Best fit parameter~$\mathcal{L}^{\rm vac}_\perp$ for the Poynting flux $L(\chi)/L_{\rm dip}^{\rm vac} = \mathcal{L}^{\rm vac}_\perp\,\sin^2\chi$ of the vacuum oblique rotator in Newtonian and general-relativistic case for weak and strong magnetic fields.}
\label{tab:FitFluxVacuum}
\end{table}

\section{FFE results}
\label{sec:ResultatsFFE}

To be as exhaustive as possible we proof that our conclusions extend to the plasma filled magnetosphere. Thus we undertook simulations for FFE electrodynamics including GR and QED corrections. The results are discussed in the following lines.

\subsection{FFE monopole}

\begin{figure}
\centering
\input{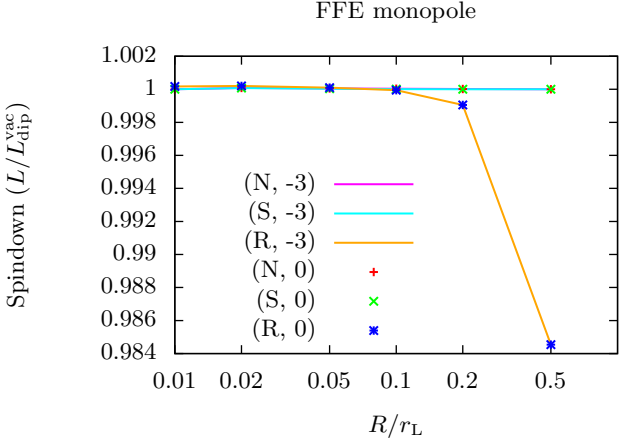}
\caption{Spindown luminosity for the FFE monopole for different rotation rates and magnetic field strengths given by the couple $(a,\log(b))$ and gravitational field (Newtonian, Schwarzschild or slowly Rotating star) as indicated in the legend.}
\label{fig:Spindown_Monopole_FFE}
\end{figure}

The FFE monopole case is of particular importance because exact analytical solutions are known in special relativity although a monopole magnetic field is not realistic. We compare the spindown luminosity found from our simulations to the exact analytical expression in classical Newtonian gravity and in general relativity. The analytical results is retrieved to good accuracy in the classical but also in the quantum case. Actually both results are undistinguishable, they overlap nicely, see fig.~\ref{fig:Spindown_Monopole_FFE}. We conclude that QED has little impact on the FFE monopole structure and can be neglected. QED corrections do not have a significant impact on the global topology of the plasma filled magnetosphere.

\subsection{FFE aligned dipole}

\begin{figure}
\centering
\input{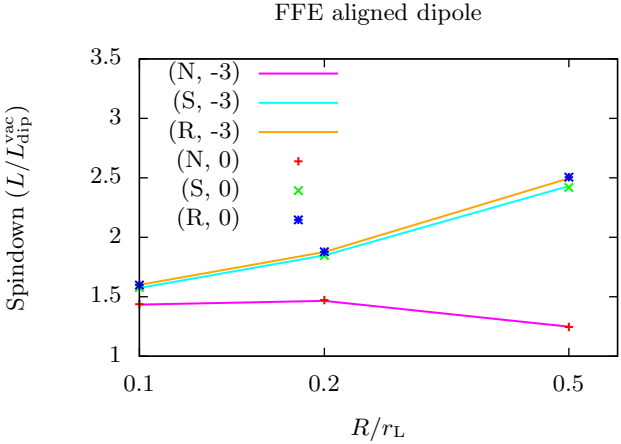}
\caption{Spindown luminosity for the FFE aligned rotator for different rotation rates and magnetic field strengths given by the couple $(a,\log(b))$ and gravitational field (Newtonian, Schwarzschild or slowly Rotating star) as indicated in the legend.}
\label{fig:Spindown_Dipole_Aligne_FFE}
\end{figure}

The same study is performed for the rotating dipole. We compare again the spindown luminosity obtained from several simulations. Classical and quantum cases are here also undistinguishable, see fig.~\ref{fig:Spindown_Dipole_Aligne_FFE}. QED has little impact on the FFE dipole structure and can be neglected. 
\begin{table}
\centering
\begin{center}
\begin{tabular}{ccccc}
\hline 
 & \multicolumn{2}{c}{Newtonian} & \multicolumn{2}{c}{General-relativistic} \\
\hline 
\diagbox{$a$}{$\log b$} & $-3$ & $0$ & $-3$ & $0$ \\
\hline 
0.1  & 1.4342 & 1.4373 & 1.5996 & 1.5988 \\
0.2  & 1.4649 & 1.4722 & 1.8759 & 1.8778 \\
0.5  & 1.2488 & 1.2455 & 2.4950 & 2.5051 \\
\hline
\end{tabular}
\end{center}
\caption{Spindown luminosity for the FFE aligned rotator in several approximations.}
\label{tab:Spindown_FFE_Aligne}
\end{table}

\subsection{FFE orthogonal dipole}

\begin{figure}
\centering
\input{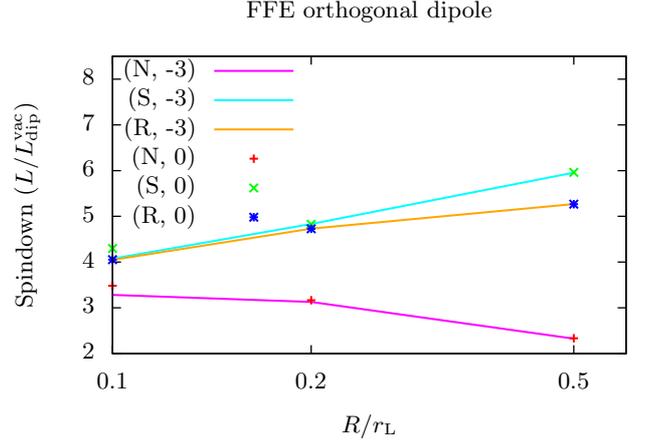}
\caption{Spindown luminosity for the FFE orthogonal rotator for different rotation rates and magnetic field strengths given by the couple $(a,\log(b))$ and gravitational field (Newtonian, Schwarzschild or slowly Rotating star) as indicated in the legend.}
\label{fig:Spindown_Dipole_Perp_FFE}
\end{figure}

Let us continue with the most interesting case, the orthogonal rotator. We also investigated the influence of the neutron spin depicted by the ratio $R/\rlight$. Results are shown in fig.~\ref{fig:Spindown_Dipole_Perp_FFE}. There is no distinction between classical and QED spindown luminosity. Both cases overlap again to high accuracy.
\begin{table}
\centering
\begin{center}
\begin{tabular}{ccccc}
\hline 
& \multicolumn{2}{c}{Newtonian} & \multicolumn{2}{c}{General-relativistic} \\
\hline 
\diagbox{$a$}{$\log b$} & $-3$ & $0$ & $-3$ & $0$ \\
\hline 
0.1  & 3.2829 & 3.4827 & 4.0464 & 4.0526 \\
0.2  & 3.1279 & 3.1661 & 4.7310 & 4.7267 \\
0.5  & 2.3255 & 2.3341 & 5.2681 & 5.2670 \\
\hline
\end{tabular}
\end{center}
\caption{Spindown luminosity for the FFE orthogonal rotator in several approximations.}
\label{tab:Spindown_FFE_Perp}
\end{table}

\subsection{FFE oblique dipole}

We finish with the oblique rotator by estimating the dependence of the spindown luminosity on the inclination angle~$\chi$ of the dipole. The Poynting flux with respect to $a$, $\log(b)$ and $\chi$ is shown in fig.~\ref{fig:Spindown_Dipole_Oblique_N_FFE} for Newtonian gravity and in fig.~\ref{fig:Spindown_Dipole_Oblique_GR_FFE} for general relativity. Here again we do not notice any significant deviation from the classical approximation neither in the Newtonian regime nor in general relativity.
\begin{figure}
\centering
\input{oblique_m2_j1_o2_K64_N32_cfl5_ro10_g0.tex}
\caption{Spindown luminosity for oblique rotators for different rotation rates and magnetic field strengths given by the couple $(a,\log(b))$ as indicated in the legend for Newtonian gravity. Red solid lines are best fits.}
\label{fig:Spindown_Dipole_Oblique_N_FFE}
\end{figure}
\begin{figure}
\centering
\input{oblique_m2_j1_o2_K64_N32_cfl5_ro10_g2.tex}
\caption{Spindown luminosity for oblique rotators for different rotation rates and magnetic field strengths given by the couple $(a,\log(b))$ as indicated in the legend in general relativity. Red solid lines are best fits.}
\label{fig:Spindown_Dipole_Oblique_GR_FFE}
\end{figure}
The points are taken from the simulations whereas the solid curves are best fits obtained by adjusting to a $\sin^2\chi$ dependence such that
\begin{equation}
 \frac{L}{L_{\rm dip}^{\rm FFE}} = \mathcal{L}^{\rm FFE}_\parallel + \mathcal{L}^{\rm FFE}_\perp \, \sin^2\chi .
\end{equation} 
The precise values are reported in Table~\ref{tab:FitFluxFFE} comparing Newtonian and general-relativistic gravity. As the ratio~$R/\rlight$ decreases, both kind of curves, Newtonian and general-relativistic, tend to the function~$\sin^2\chi$, the former from below and the latter from above as was already noticed for the vacuum case. Note that for all dipolar FFE magnetosphere with~$a=0.1$, the discrepancy between classical and quantum results deviate more than expected because of the numerical resolution which should be increased to accurately resolve the polar caps. It is an effect of the grid not physics.
\begin{table*}
\centering
\begin{center}
\begin{tabular}{ccccc}
\hline
 & \multicolumn{2}{c}{Newtonian} & \multicolumn{2}{c}{GR} \\
 \hline
\diagbox{$a$}{$\log b$} & $-3$ & $0$ & $-3$ & $0$ \\
\hline
\hline
0.1 & (1.4428,1.8527) & (1.4479,2.0037) & (1.5890,2.4515) & (1.6232,2.3308) \\
0.2 & (1.4628,1.6818) & (1.4709,1.7112) & (1.8785,2.8682) & (1.8742,2.8873) \\
0.5 & (1.2597,1.0896) & (1.2614,1.1006) & (2.5453,2.7903) & (2.5515,2.7870) \\
\hline
\end{tabular}
\end{center}
\caption{Best fit parameters~$(\mathcal{L}^{\rm FFE}_\parallel,\mathcal{L}^{\rm FFE}_\perp)$ for the Poynting flux $L(\chi)/L_{\rm dip}^{\rm FFE} = \mathcal{L}^{\rm FFE}_\parallel + \mathcal{L}^{\rm FFE}_\perp \, \sin^2\chi$ of the FFE oblique rotator in Newtonian and general-relativistic case for weak and strong magnetic fields.}
\label{tab:FitFluxFFE}
\end{table*}

\section{Conclusion}
\label{sec:Conclusion}

Strongly magnetized rotating fields in vacuum such as the one expect in neutron star systems and especially in magnetars are responsible for their electromagnetic activity like pair creation and very high energy emission processes which are effectively observed at Earth. Thus in some sense we get indirect insights into the physics of such strong fields. In this paper we have shown that despite the presence of magnetic field strengths around the critical field, QED corrections would not lead to drastic changes in the global electrodynamics of a neutron star magnetosphere, especially not in the rate of braking through electromagnetic radiation of the large amplitude low frequency electromagnetic wave in vacuum. Filling the magnetosphere with a high density ultra-relativistic pair plasma leading to the force-free regime does not modify this outcome.

So far our results have been restricted to fields $B\lesssim\BQ$ because of the first order QED Lagrangian we used. However, for very intense fields $B\gg\BQ$ we do not expect the QED effect to become dominant because asymptotically, the perturbation of the Lagrangian scales as $\ln (B/\BQ)$ and would require unrealistically high fields to become comparable to the unperturbed Lagrangian \citep{LandauLifchitzTome4}. Consequently, our results are fairly robust even in the extreme case of high-B field magnetars.


\section*{Acknowledgements}

I am very grateful to the referee for his valuable comments and suggestions. This work has been supported by the French National Research Agency (ANR) through the grant No. ANR-13-JS05-0003-01 (project EMPERE). It also benefited from the computational facilities available at Equip@Meso from Universit\'e de Strasbourg.


\end{document}